\documentclass[prd,aps,superscriptaddress,twocolumn,floatfix,nofootinbib, preprintnumbers]{revtex4-2}
\pdfoutput=1

\usepackage{amsfonts}
\usepackage{amsmath}
\usepackage{amssymb}
\usepackage{bm}
\usepackage{dcolumn}
\usepackage{graphicx}   
\usepackage[latin1]{inputenc}
\usepackage{latexsym}
\usepackage{rotating}
\usepackage{hyperref}
\usepackage{graphicx}
\usepackage{color}
\usepackage{mathrsfs}
\usepackage{calc}
\usepackage{braket}

\newcommand{\pd}{{\rm d}}
\newcommand{\p}{\partial}
\newcommand{\tr}{{\rm tr}}
\newcommand{\ceq}{\stackrel{\circ}{=}}
\newcommand{\cceq}{\stackrel{\circ\circ}{=}}
\newcommand{\sfA}{\mathsf{A}}
\newcommand{\sfB}{\mathsf{B}}

\newcommand\be{\begin{equation}}
\newcommand\ba{\begin{eqnarray}}
\newcommand\ee{\end{equation}}
\newcommand\ea{\end{eqnarray}}
\begin{document}

\preprint{MS-TP-24-07}

\title{Embedded Domain Walls and Electroweak Baryogenesis}

 \author{Tobias Schr{\"o}der}
   \email{schroeder.tobias@uni-muenster.de}
  \affiliation{Institute for Theoretical Physics, University of M{\"u}nster, 48149 M{\"u}nster, Germany}
 \author{Robert Brandenberger}
\email{rhb@physics.mcgill.ca}
\affiliation{Department of Physics, McGill University, Montr\'{e}al,
  QC, H3A 2T8, Canada}


\begin{abstract}

Embedded walls are domain wall solutions which are unstable in the vacuum but stabilized in a plasma of the early Universe.  We show how embedded walls in which the electroweak symmetry is restored can lead to an efficient scenario of electroweak baryogenesis.  We construct an extension of the Standard Model of particle physics in which embedded walls exist and are stabilized in an electromagnetic plasma.

\end{abstract}

\maketitle

\section{Introduction} 
\label{sec:intro}

 Electroweak baryogenesis is an interesting scenario to explain the origin of the observed asymmetry between matter and antimatter (see, e.g., \cite{EWBGrevs} for some reviews). As realized a long time ago \cite{Sakharov}, in order to be able to generate an asymmetry between baryons and antibaryons starting from symmetric initial conditions, three criteria must be satisfied. First, the theory needs to admit baryon number violating processes. Secondly, the CP symmetry must be broken in the sector of the theory which communicates with baryons, and thirdly, the baryon number violating processes must take place out of thermal equilibrium.  There is CP violation in the Standard Model of particle physics, and, as realized in \cite{KRS}, the Standard Model also features baryon number violating sphaleron processes. The key challenge is to realize a setup in which the baryon number violating processes take place out of thermal equilibrium. 
 
 If the electroweak phase transition is strongly first order, then it proceeds by the nucleation and growth of bubbles of the broken phase in a sea of the unbroken phase. The bubble walls represent regions of space-time which are out of thermal equilibrium. Hence, sphaleron processes which take place inside of the bubble walls can lead to baryogenesis \cite{EWBGrevs}.  However, in the Standard Model, the electroweak phase transition is not strongly first order, and thus, the above mechanism is ineffective.  Going beyond the Standard Model, it is possible to construct models in which a strong first-order electroweak phase transition is realized.
 
 However, as pointed out in \cite{Trodden, Prokopec}, there is another way to obtain regions of space-time which are out of thermal equilibrium, namely by invoking topological defects. It is known that in large classes of particle physics models beyond the Standard Model, topological defect solutions exist (see, e.g., \cite{VS, HK, RHBreview} for reviews of the role of topological defects in cosmology). If Nature is described by a theory with defect solutions, then causality implies that a network of defects will form in the early universe \cite{Kibble}. If the defects are topologically stable, the network of defects will persist to all times \cite{Kibble}. Topological defects are out-of-equilibrium configurations. Thus, provided the electroweak symmetry is unbroken inside of the defects, the defects can be the locations in space-time where electroweak baryogenesis takes place.  Among theories with topologically stable defects, those with cosmic strings are the most interesting since, in this case, the string network contributes a fixed fraction of the total energy density. However, it was found in \cite{Prokopec} (see also \cite{Moore}) that in the case of cosmic strings, baryogenesis occurs in too small a volume of space and is unable to generate the observed net baryon-to-entropy ratio.  As was already remarked in \cite{Prokopec}, if the defects were domain walls, baryogenesis could take place in all of space and hence be effective. But models with stable domain walls are ruled out since a single domain wall would overclose the universe \cite{DWproblem} \footnote{Assuming here that the energy scale of the domain walls is larger than the LHC scale.}. As we point out here,  ``embedded walls'' can help to provide an efficient mechanism of electroweak baryogenesis \footnote{Yet another baryogenesis mechanism via electroweak-symmetric balls was suggested in \cite{Bai}.}.
 
 An ``embedded defect'' is defined as a defect configuration which is not topologically stable in the vacuum, but can be stabilized by plasma effects. A simple example is the ``electroweak Z-string'', a string configuration in the standard electroweak model which can be stabilized in an electromagnetic plasma \cite{Nagasawa} \footnote{See also \cite{Ana} for other types of non-topological defects, and \cite{Karouby} for further work on the plasma stabilization mechanism.}. The Higgs field of the electroweak theory is a complex Higgs doublet, one of the complex fields electrically charged and the other one neutral. Through the gauge kinetic term, in a plasma, the vacuum manifold (which is $S^3$ without plasma interactions) is lifted in the charged field direction. The remaining effective vacuum manifold is $S^1$, thus allowing for cosmic string solutions in which the charged scalar field is zero, and the neutral one winds the vacuum manifold. Similarly, the low energy effective theory of the strong interactions has two complex scalar fields, one of them charged (the charged pions) and the other neutral (representing the neutral pion and the sigma field). In the case of vanishing quark masses, the vacuum manifold is $S^3$, but coupling to an electromagnetic plasma lifts the potential in the charged pion directions, leaving the vacuum manifold of the effective theory to be $S^1$. In this case, the ``pion string'' \cite{pionstring} is stabilized and can, in fact, be used to generate primordial magnetic fields \cite{Zhang} \footnote{The plasma stabilization mechanism of embedded strings has many potential phenomenological applications beyond the utilization of pion strings for magnetogenesis. This is a rich area of research which we plan to further explore in the future.}.
 
 In this paper we will first show that embedded domain walls which are stabilized in the electromagnetic plasma of the early universe can lead to an efficient scenario of electroweak baryogenesis, provided that the electroweak symmetry is unbroken in the core of the walls. We then present an extension of the Standard Model of particle physics where embedded walls with the required properties arise.
 
 We begin with a short review of plasma stabilization of embedded defects. Section (\ref{section3}) is an analysis of wall-mediated electroweak baryogenesis, and in Section (\ref{section4}), we estimate the net baryon-to-entropy ratio, which can be obtained from embedded walls. In Section (\ref{section5}), we present a particle physics model in which embedded walls with the required properties arise. In the final section, we discuss our results.
 
We work in the context of a spatially flat space-time with metric
\be
ds^2 \, = \, dt^2 - a^2(t) d{\bf{x}}^2 \, ,
\ee
where $t$ is physical time, ${\bf{x}}$ are the spatial comoving coordinates, $a(t)$ is the scale factor, and we use natural units in which $c = \hbar = k_B = 1$. The temperature is denoted by $T$. The baryogenesis processes we are interested in take place in the radiation phase of cosmology,  and $g^*$ will denote the effective number of entropic degrees of freedom present at the relevant time.

\section{Embedded Defects - a Brief Review}
\label{section2}
 
We will illustrate the idea behind embedded defects with the example of the electroweak Z-string, an embedded string in the standard electroweak theory. The Higgs field is a complex Higgs doublet. In terms of the four real component fields $\phi_i: i = 0,1,2,3$ the potential is
 \be \label{barepot}
 V(\phi) \, = \, \frac{1}{4} \lambda \left( \sum_{i=0}^{3} \phi_i^2  - \eta^2 \right)^2 \, ,
 \ee
 where $\lambda$ is the Higgs coupling constant and $\eta$ is the vacuum expectation value of the field in the broken phase.  The fields $\phi_0$ and $\phi_3$ are uncharged (under the usual $U(1)$ of electromagnetism), while $\phi_1$ and $\phi_2$ are charged. The vacuum manifold is $S^3$ and hence there are no stable topological defects (in four space-time dimensions).  The Lagrangian of the scalar sector of the Standard Model after electroweak symmetry breaking is
 \be
 {\cal{L}} \, = \, \frac{1}{2} \mathcal{D}^{\mu} \phi_i \mathcal{D}_{\mu} \phi_i - V(\phi) - \frac{1}{4} F^{\mu \nu} F_{\mu \nu} \, ,
 \ee
 where the index $i$ runs from 0 to 3, and $\mathcal{D}_{\mu} = \partial_{\mu} - i q_i A_{\mu}$ is the gauge covariant derivative operator,  $q_i$ being the charge of the $\phi_i$ field.  $F_{\mu \nu}$ is the field strength tensor associated with the gauge fields $A_{\mu}$. 
 
 It is well known that if $\phi$ is in thermal equilibrium, then at high temperature, the full gauge symmetry can be restored since $\phi = 0$ is the minimum of the finite temperature effective potential \cite{finiteT} \footnote{
Note that, as already pointed out in \cite{W}, there are models in which the electroweak symmetry is not restored (see e.g. \cite{L1, L2} for some lattice models, and \cite{M} for a more recent analysis), or restored only at temperatures much higher than the electroweak symmetry breaking scale (see e.g. \cite{Ser}).}. One way to understand this symmetry restoration is to study the effect of thermal fluctuations of the fields (scalars, vector and spinors) on a scalar field background. Due to the nonlinearities in the Lagrangian, the fluctuations of $\phi$ contribute a correction term (see, e.g., \cite{RMP} for a review of finite temperature effects on the effective potential)
 \be
 \delta V \, \sim \, \lambda T^2 \phi^2 \, ,
 \ee
 where $T$ is the temperature. Gauge field fluctuations contribute a similar term, but with $\lambda$ replaced by the gauge coupling constant. At temperatures larger than a critical value $T_c$, $\phi = 0$ becomes the minimum of the effective potential.
 
When discussing topological defect formation, one is interested in the situation when $\phi$ is no longer in thermal equilibrium, but one of the gauge fields is.  In the case of the Standard Model, after electroweak symmetry breaking, the photon field is the only field which remains massless below the symmetry breaking scale and which is then in thermal equilibrium. In this case, one-loop photon effects from the gauge kinetic term produce \cite{Nagasawa} a contribution to the effective potential which lifts the potential, but only in the charged scalar field directions, i.e.
\be
\delta V \, \sim \, g T^2 \left( \phi_1^2 + \phi_2^2 \right) \, .
\ee
This effect reduces the vacuum manifold to $S^1$
\be
{\cal{M}}_{\rm eff} \, = \, \left\{\left(\phi_0, \phi_3\right): \phi_0^2 + \phi_3^2 = \eta^2 \right\} \, .
\ee
This allows for string solutions, solutions where the neutral scalar field winds ${\cal{M}}_{\rm eff}$.

In order to obtain embedded walls, we must have a theory in which plasma effects break a gauge symmetry to a discrete symmetry such that ${\cal{M}}_{\rm eff}$ is disconnected. A toy model would be a scalar field triplet $\phi = \left(\phi_0, \phi_1, \phi_2\right)$ for which $\phi_1$ and $\phi_2$ are electrically charged while $\phi_0$ is neutral. If the bare potential for $\phi$ is of the usual symmetry breaking form (\ref{barepot}) (with the sum over $i$ now running from $0$ to $2$), then in an electromagnetic plasma the effective potential will be
\be
{\cal{M}}_{\rm eff} \, = \, \left\{ \left(\phi_0, \phi_1, \phi_2\right) = \left(\pm \eta, 0, 0\right) \right\} \,
\ee
and will hence allow for embedded domain walls, walls across which $\phi_0$ transits from $\phi_0 = \eta$ to $\phi_0 = - \eta$.

In Section (\ref{section5}) we will present an extension of the Standard Model in which embedded walls can be realized,  inside of which the electroweak symmetry is restored. First, we will discuss how such embedded walls can lead to an effective scenario of electroweak baryogenesis.

\section{Electroweak Baryogenesis from Embedded Walls}
\label{section3}

We consider a scenario which yields embedded domain walls inside of which the electroweak symmetry is unbroken. Baryon number violating processes are unsuppressed inside of the domain wall, and the wall boundaries represent the location in which the out-of-equilibrium condition is satisfied. There are two general scenarios of electroweak baryogenesis - local \cite{localBG} and nonlocal \cite{CKN, Joyce1, Joyce2, Joyce3}.  In local baryogenesis, baryon number violation takes place in the same region of space where the out-of-thermal-equilibrium condition is satisfied, namely in the boundary region.  In nonlocal baryogenesis, scattering of particles off the advancing wall edge produces a chiral fermion current, which flows to the inside of the domain wall, where it transforms to a net baryon number via sphaleron processes. 

Before we turn more closely to the case of non-local baryogenesis, we take a brief look at local baryogenesis to see that it is insufficient to reproduce the observed baryon-to-entropy ratio. Consider a point in space ${\bf{x}}$ which is approached by a domain wall. A net anti-baryon number is generated in the leading edge. Weak sphaleron processes then lead to a relaxation of this number while ${\bf{x}}$ is inside the wall where sphaleron processes are unsuppressed.  A baryon number with an opposite sign to what is generated when the leading wall edge passes ${\bf{x}}$ is then generated when the trailing edge passes ${\bf{x}}$ and is not relaxed since it stays outside the domain wall, leading to a net positive baryon number. The necessary CP violation on the domain wall can, for example, be achieved by introducing a second Higgs doublet where the CP violation occurs from a relative phase $\theta$ between the two Higgs fields. In this case, the baryon-to-entropy ratio produced locally at the edge of the wall was estimated in \cite{Prokopec} to be
\begin{align}
    \frac{\tilde{n}_b^0}{s} \simeq -4 \frac{\Gamma_s}{g^* T^4}\left(\frac{m_f}{T}\right)^2 \theta_{CP} \label{eq:localedgeBG}
\end{align}
where the entropy density reads
\be
s = \frac{2\pi^2}{45}g^* T^3. \, \label{eq:entropydensity}
\ee
Here, $\theta_{CP}$ denotes the change in the CP-violating relative angle in the two-Higgs-doublet model over the domain wall boundary, and $\Gamma_s$ is the weak sphaleron rate  per unit volume \cite{Ambjorn}
\be
\Gamma_s = \kappa \alpha_W^4 T^4 \, ,
\ee
with $\alpha_W = \frac{g^2}{4\pi}$, and a constant $\kappa$ in the range $\kappa\sim 0.1\dots 1$ \footnote{In \cite{Boedeker} a diffent dependence on the electroweak coupling $\Gamma_s \propto \ln\left(\frac{1}{\alpha_W}\right) \alpha_W^5 T^4$ was found. For the numerical evaluation, this difference is, however, covered by the prefactor $\kappa$.}. In principle, both quarks and leptons, with $m_f$ denoting their mass, can contribute to this process. However, in the case of quarks, any induced CP-violation in the domain wall boundary that could be converted via weak sphalerons will be washed out by the much faster strong sphalerons. Therefore, only leptons are expected to have a significant impact. From the scaling of \eqref{eq:localedgeBG} with the fermion mass, it is clear that the most massive lepton, i.e. the tau, will dominate the baryon production. 
Let us consider for definiteness $\theta_{CP}$ to be positive such that a net antibaryon number is produced at the leading edge of the domain wall. Since these antibaryons still have to pass through the region of width $l_D$ in which electroweak symmetry is restored, weak-sphalerons will wash out the baryon number again, and their number density is suppressed by a factor of $e^{-\frac{\bar{\Gamma}_s l_D}{v_D}}$ after the passage. Here, $\bar{\Gamma}_s= 6N_f\frac{\Gamma_s}{T^3}$ is the decay rate of the antibaryons, and $v_D$ is the velocity of the domain wall such that the passage time is $l_D/v_D$. $N_f$ denotes the number of particle families, and we will use $N_f=3$ for all numerical evaluations.

Since the baryons produced at the trailing edge do not pass through the domain wall, they are not diluted, and the net baryon-to-entropy ratio after the passage of the domain wall becomes
\begin{align}
    B\equiv \frac{n_b}{s} = \frac{\tilde{n}_b^0}{s}\left(e^{-\frac{\bar{\Gamma}_s l_D}{v_D}} - 1 \right) \sim 10^{-16}.
\end{align}
For the numerical value, we took $l_D\sim m_H^{-1}$, $\kappa, \theta_{CP}, v_D \sim 1$, $g^*\sim 10^2$, and $T\sim m_H$.
This value is at least $5$ orders of magnitude too small to explain the observed baryon-to-entropy ratio, even though we considered the parameters in a range in which it becomes maximal. 

Therefore, let us now turn to non-local electroweak baryogenesis \cite{CKN, Joyce1, Joyce2, Joyce3} \footnote{
The calculation of the CP-violating source is complex, with many possible contributions (see, e.g. \cite{EWBG1, EWBG2} for more recent reviews). Our analysis is based on calculating reflection and transmission coefficients at the wall's boundary and gives only a rough order of magnitude estimate. For a recent discussion of some of the subtle aspects, see \cite{Ramsey}.}. In this mechanism, after the electroweak phase transition, a net chiral fermion number builds up at ${\bf{x}}$ as the leading wall edge passes the point and afterwards, this chiral current diffuses into the wall. Since, by assumption, the electroweak symmetry remains unbroken inside the wall, weak sphaleron processes lead to net baryon number generation while ${\bf{x}}$ is inside the wall.

Let us now be more quantitative. Generally, for $N_F$ families,  the baryon production rate (per unit volume) is given by
\begin{align}
    \frac{\pd n_b}{\pd t} =  - \frac{N_F\Gamma_s}{2T} \sum_i \left(3\mu_{u_L}^i + 3\mu_{d_L}^i + \mu_{l_L}^i + \mu_{\nu_L}^i\right) \label{eq:baryonnumber1}
\end{align}
where the $\mu$ denote differences between particle and antiparticle chemical potentials, the index $i$ runs over the families, $u,d,l,\nu$ denoting up-, down-type quarks, charged leptons and neutrinos, respectively. The index $L$ denotes left-handedness.  
 
 For massless fermions, chemical potentials and perturbations in the associated number density $\delta n$ are related via 
 \be
 \delta n = \frac{\mu T^2}{12}
 \ee
 such that equation \eqref{eq:baryonnumber1} may be rewritten at sufficiently large temperatures as
\begin{align}
    \frac{\pd n_b}{\pd t} = -\frac{6N_F \Gamma_s}{T^3}\left(3n_{b, L} + n_{l,L}\right) \label{eq:baryonnumberconversionrate}
\end{align}
where $n_{b,L}, n_{l,L}$ denote the total number densities of left-handed baryons and leptons.  

The CP violation on the domain wall due to the relative phase $\theta$ between the two Higgs fields in the two-Higgs-doublet model leads, for sufficiently thin walls as will be realized in our scenario, to differences between reflection coefficients of particles and anti-particles on the domain wall. 

In the following, we fix our coordinates such that the domain wall lies in the $xy$-plane at $z = 0$. Let us denote the reflection coefficient for right-handed fermions incident from the unbroken phase by $R_{R\to L}^u$ and the transmission coefficient from the unbroken to the broken phase by $T_R^{u\to b}$.  Furthermore, denoting by $\bar{L}$ and $\bar{R}$ the CP-conjugates of left- and right-handed particles, respectively, it was found that the difference satisfies \cite{Joyce2}
\begin{align}
    \Delta R(p_z)&= R^u_{R\to L} - R^u_{\bar{R}\to \bar{L}} = \\
    &= 4 t(1-t^2) \theta_{CP} \frac{m_f}{m_H} \exp\left(-\frac{p_z^2}{m_H^2}\right)\Theta(|p_z|-m_f),\nonumber
\end{align}
where $t = \tanh\left(\vartheta\right)$, $\tanh\left(2\vartheta\right) \equiv \frac{m_f}{|p_z|}$, $\theta_{CP}$ is the change in the CP-violating angle over the wall's boundary, $m_H$ is the mass of the electroweak Higgs, $m_f$ is the mass of the fermion in the broken phase and $p_z$ the $z$-component of its momentum in the wall frame. The $\Theta$-function occurs due to the fact that particles with $|p_z| < m_f$ will be totally reflected from the wall and consequently $\Delta R = 0$, while the suppression of large momenta $|p_z| > m_H$ is due to coherence effects across the boundary of the region in which the electroweak symmetry is restored. The CP-violating angle changes over the width of this boundary, which we assumed to be $\sim m_H^{-1}$. Since $|p_z|>m_f$, we can approximate $t\simeq \frac{m_f}{2|p_z|}$ and, furthermore, replace the exponential by a step function, thus obtaining
\begin{align}
    \Delta R(p_z) \simeq  \frac{2m_f^2}{|p_z|m_H} \theta_{CP} \Theta\left(m_H-|p_z|\right) \Theta\left(|p_z|-m_f\right). \label{eq:DeltaRsimplified}
\end{align}

Next, we want to consider the net flux of left-handed particles $J_0$ into the unbroken phase. For this, we need to calculate the difference between the flux densities $j_L$ and $j_{\bar{L}}$ of left-handed particles and their CP conjugates such that
\be
J_0 = \int \frac{\pd^3 p}{(2\pi)^3} \left(j_L-j_{\bar{L}}\right)\, .
\ee
For both $j_i$, we have to consider transmission from the broken into the unbroken phase and reflection back into the unbroken phase \cite{CKN}. For concreteness, let us assume that the domain wall moves with velocity $v_D$ in positive $z$-direction. We want to calculate the influx into the domain wall from the broken phase on the right into the unbroken phase inside the domain wall on the left. We can then write, e.g., 
\begin{align}
j_L(p_z, p_\bot) =& f^{\leftarrow}(p_z, p_\bot) T^{b\to u}_L+f^{\rightarrow}(p_z, p_\bot) R_{R\to L}^u -\nonumber\\ &-  f^{\rightarrow}(p_z, p_\bot)\, , \label{lcurrent}
\end{align}
where $f^\rightarrow$ denotes the flux density of particles moving in the unbroken phase to the right (towards the domain wall boundary) and $f^\leftarrow$  that of particles moving in the broken phase to the left (also towards the domain wall boundary). Here, $p_z$ is the momentum perpendicular to the domain wall and $p_\bot=\sqrt{p_x^2+p_y^2}$. For the third term in \eqref{lcurrent} we used that $R_{L\to R}^u+T^{u\to b}_L = 1$ which implies that this term will cancel when considering $j_L-j_{\bar{L}}$.

Assuming free field phase space densities and using the velocity $v_z = \frac{p_z}{E}$, we have in the wall frame
\be
    f^\leftarrow = \frac{|p_z|}{E} \frac{1}{1+\exp\left(\frac{\gamma}{T}\left(E-v_D\sqrt{p_z^2-m_f^2}\right)\right)} 
\ee
and
\be
f^\rightarrow = \frac{|p_z|}{E} \frac{1}{1+\exp\left(\frac{\gamma}{T}\left(E+v_D|p_z|\right)\right)}
\ee
where $E= \sqrt{p_\bot^2+p_z^2}$ is the energy in the wall frame and $\gamma = \frac{1}{\sqrt{1-v_D^2}}$. 

We can now relate transmission and reflection coefficients by first using CPT invariance, then probability conservation, and again CPT invariance
\begin{align}
    T^{b\to u}_L = T^{u\to b}_{\bar{L}} = 1- R^u_{\bar{L}\to \bar{R}} = 1- R^u_{R\to L}.
\end{align}
Doing the same for $j_{\bar{L}}$, one finds the expression 
\ba
    J_0 &=&  \int_{p_z < 0} \frac{d^3 p}{\left(2\pi\right)^3}  \left(f^\rightarrow - f^\leftarrow\right) \Delta R(p_z) \\
    &=& \frac{m_f^2}{2\pi^2 m_H}\theta_{CP}\int_{m_f}^{m_H} \pd p_z \int_0^\infty p_\bot \pd p_\bot \frac{f^\rightarrow-f^\leftarrow}{|p_z|} \, . \nonumber \label{eq:generalexpressionlefthandedflux}
\ea
First, let us assume that we are at temperatures for which $\frac{m_H}{T} \ll 1$ and use $\frac{m_f}{m_H}\ll 1$. We then obtain \cite{Joyce2}
\begin{align}
    J_0 = \frac{v_D m_f^2 m_H \theta_{CP}}{4\pi^2}.
\end{align}

In a similar manner, we can calculate the average velocity of the chiral flux relative to the wall \cite{Joyce2}
\begin{align}
    v_i \equiv \frac{\int_{p_z<0} \frac{\pd^3 p}{(2\pi)^3} \frac{|p_z|}{E} \left(f^\rightarrow-f^\leftarrow\right) \Delta R(p_z)}{\int_{p_z<0} \frac{\pd^3 p}{(2\pi)^3}\left(f^\rightarrow-f^\leftarrow\right) \Delta R(p_z)} \simeq \frac{1}{4\ln\left(2\right)}\frac{m_H}{T} \label{particlevelocities}
\end{align}
where we expanded in leading order of $v_D, \frac{m_f}{m_H}, \frac{m_H}{T}\ll 1$ as before.

Next, in order to compute the conversion of the injected chiral fermion current inside the wall, we have to consider the diffusion equation. Modelling the injected current as $\xi_p J_0 \delta\left(z-v_Dt\right)$, the first and second diffusion laws together with $\dot{n}_{l,L} = -v_D n_{l,L}'$ can be brought into the form
\begin{align}
    D n_{l,L}'' + v_D n_{l,L}' = \xi_pJ_0 \delta'(z-v_D t).   
\end{align}
where $D= 1 / (8\alpha_W^2 T)$ is the diffusion constant, $\xi_p$ is the persistence length, and a prime denotes a derivative with respect to $z$. The persistence length can be estimated as $\xi_p \sim 6 D v_i$ \cite{Joyce2}. This equation is solved by \cite{Prokopec}
\begin{align}
    n_{l,L}(z) = J_0 \frac{\xi_p}{D}e^{-\frac{v_D}{D}z}.
\end{align}

Finally, making use of \eqref{eq:baryonnumberconversionrate} and $\dot{n}_b = -v_D n_b'$, we find
\begin{align}
    n_b' = \frac{6N_F\Gamma_s}{v_D T^3}n_{l,L} = \frac{\Bar{\Gamma}_s}{v_D} n_{l,L}
\end{align}
with 
\be
\Bar{\Gamma}_s = 6N_F \kappa \alpha_W^4 T\, . 
\ee
Integrating over the region of non-vanishing sphaleron transitions, i.e., from $0$ to $l_D$, we obtain
\begin{align}
    n_b = \underbrace{\frac{\bar{\Gamma}_s}{4\pi^2}\frac{D}{v_D} \frac{\xi_p}{D} \theta_{CP} m_f^2 m_H}_{\equiv n_b^0} \left(1-e^{-\frac{v_D}{D}l_D}\right) \label{eq:baryontoentropyratio}
\end{align}
for the induced baryon number density.
 
Using the expression \eqref{eq:entropydensity} for the entropy density, we find the following result for the prefactor of the induced baryon-to-entropy ratio in \eqref{eq:baryontoentropyratio}:
\begin{align} \label{result}
    \frac{n_b^0}{s}= \frac{45}{8\pi^4 g^*} \frac{\bar{\Gamma}_s D}{v_D} \frac{\xi_p}{D}\theta_{CP} \left(\frac{m_f}{T}\right)^2 \frac{m_H}{T}. 
\end{align}
Note that this holds only as long as $\frac{m_H}{T}\ll 1$. Applying the above formula to lower temperatures would suggest a significant rise of $n_b/s$ at late times. However, solving all the integrals numerically shows that this behaviour is not realized and, instead, that (in the relevant temperature range with $m_\tau > T$) $n_b/s$ is a slowly decreasing function of $T$ \footnote{In particular, using $\frac{\xi_p}{D}= 6v_i$, we cannot apply \eqref{particlevelocities} as this velocity becomes quickly larger than $1$ for low temperatures. }. 

While the above result can be applied to any chiral fermion current injected into the wall, quarks were found to yield negligible contributions for baryogenesis compared to left-handed leptons (see \cite{Joyce2, Vischer}). Besides quarks having shorter diffusion lengths, this is due to strong sphaleron processes in equilibrium, which quickly wash out chiral perturbations in quarks but not in leptons. The dominant contribution to this baryogenesis scenario comes, therefore, from $\tau$-leptons, on which we will focus henceforth. 

Based on (\ref{eq:baryontoentropyratio}), we can estimate the induced baryon-to-entropy ratio at a point in space which is passed once by a wall.  Since the computations above are for order of magnitude estimates still reliable at temperatures close to the value of the Higgs mass, we will use $T = m_H$ in the following.  Using $v_D^2 / (\overline{\Gamma}_s D) \sim 10^2$, $v_D \sim 1$, $\theta_{CP} \sim 1$, $v_Dl_D/D\sim 10^{-2}$, and $g^* \sim 10^2$, we find a value
\be \label{estimate}
B \, \sim \, 10^{-11} \, ,
\ee
compatible with the measured value of $B\simeq 9\times 10^{-11}$ \cite{CMB+BBN}, even if only marginally. 
This is a conservative estimate in two ways. First, it assumes that the formula (\ref{result}) ceases to apply almost immediately below the Higgs mass scale. If the formula were applicable to a lower temperature $T$, then the result would be enhanced by a factor of $(m_H / T)^3$. Secondly, the estimate (\ref{estimate}) neglects the fact that a given point of space can be passed by many wakes. In the following section, we will turn to a computation of the enhancement of baryon production due to the second effect.

 \section{Baryon-to-Entropy Ratio from Embedded Walls}
\label{section4}

Domain walls are defects in a relativistic field theory. In a theory with domain wall solutions, the system of domain walls will take on a ``scaling solution'' with a fixed number $N$ (independent of time) of walls per Hubble volume at each time and with a typical extrinsic curvature radius which is comparable to the Hubble radius. The curvature will induce motion of the walls at speeds of the order of the speed of light. Hence, if the domain wall network is sufficiently long-lived,  domain walls can swipe over each point in space multiple times. The Kibble causality argument implies that $N \geq 1$.

To obtain a rough estimate of the net baryon-to-entropy ratio induced by a network of embedded walls, we can take
\be
B^T \simeq n B \, ,
\ee
where $B$ is the result for one wall crossing from (\ref{eq:baryontoentropyratio}) and $n$ is the number of times a given point in space will be crossed by a wall during the time interval when the baryogenesis process is effective. We have
\be
n \, = {\tilde{N}} N_1 \, ,
\ee
where ${\tilde{N}}$ is the number of Hubble expansion times when baryogenesis is effective and $N_1$ is the number of wall crossings per Hubble time. The latter is given by 
\be
N_1 \, \simeq \, \frac{t v_D}{L_D} \, ,
\ee
where $L_D$ is the mean separation between walls
\be
L_D \, \sim \, \frac{t}{N} \, .
\ee
Hence,
\be
n \, \sim \, N {\tilde{N}} v_D \, .
\ee
Since embedded walls arising in quantum field theories are relativistic objects, we have $v_D \sim 1$. Numerical simulations of cosmic string evolution \cite{CSevol} indicate that a number $N \sim 10$ is reasonable. The velocity-dependent one-scale model for domain walls also yields a number $N > 1$ (see, e.g., \cite{Avelino-24} and references therein).

The number ${\tilde{N}}$ of Hubble expansion times during which electroweak baryogenesis is efficient can be estimated to be the number of expansion time steps between the time of electroweak symmetry breaking when the temperature is $T_{\rm EW}$ and the time when $T$ drops below $m_\tau$. Thus, making use of the Friedmann equation to relate time to temperature, we obtain
\be
{\tilde{N}} \, \sim \, 2 \ln\left(\frac{T_{\rm EW}}{m_\tau}\right) \, .
\ee
With $T_{\rm EW} \simeq 160 {\rm GeV}$ and $m_\tau = 1.8 {\rm GeV}$, and taking $N \sim 10$, we see that an enhancement factor of between one and two orders of magnitude over what is obtained from a single wall crossing is possible. Hence, a sufficient baryon-to-entropy ratio can be generated even if $\theta_{CP} \sim 10^{-1}$.

In the above estimate, we have neglected the decay of the baryon number produced in the $n$-th crossing when the next wall crosses, as well as the temperature dependence of the baryon production rate. An improved estimate of the total baryon-to-entropy ratio can be obtained in the following way:

For the case of local baryogenesis, we can express the baryon-to-entropy ratio after swiping over each point in space $n+1$ times as
\be 
    B_0^{n+1} \, = \, B_0^{n}\exp\left(-\frac{\overline{\Gamma}_s l_D}{v_D}\right) + \frac{\tilde{n}_b^0}{s}\left(1 - \exp\left(-\frac{\overline{\Gamma}_s l_D}{v_D}\right)\right)  \label{localBG1}
\ee
with 
\be
B_0^0 \, = \, 0 \, ,
\ee
where, as before, $l_D$ is the width of electroweak symmetry restoration around the domain wall and $\tilde{n}_b^0$ the baryon number density produced locally at its edge (cf. equation \eqref{eq:localedgeBG}).  The first factor takes into account that after the $n+1$-st passage of the domain wall, the baryons that remained after the $n$-th step decay during the passage of the wall due to sphaleron processes. The second term is the newly produced baryon number density due to the difference between baryons produced at the trailing edge and antibaryons produced at the leading edge, which had time to decay during the passage of the wall. 

For non-local baryogenesis, we have
\be
    B_0^{n+1} \, = \, B_0^{n}\exp\left(-\frac{\overline{\Gamma}_s l_D}{v_D}\right) + \frac{{n}_b^0}{s}\left(1 - \exp\left(-\frac{v_D l_D}{D}\right)\right)  \label{nonlocalBG1}
\ee
with
\be
 B_0^0 \, = \, 0 \, ,
\ee
where ${n}_b^0$ was calculated in \eqref{result}. While the first term in the previous equation remains unmodified with respect to the local case, the second term, which describes the production mechanism, is now changed as baryogenesis happens not only locally at the edge of the domain wall but in the entire domain wall due to injection of a chiral lepton current into the unbroken electroweak phase via diffusion from the domain wall edge.  Here, $D= 1 / (8\alpha_W^2 T)$ is the diffusion constant. 

\begin{figure*}
    \centering
    \includegraphics[width=0.49\linewidth]{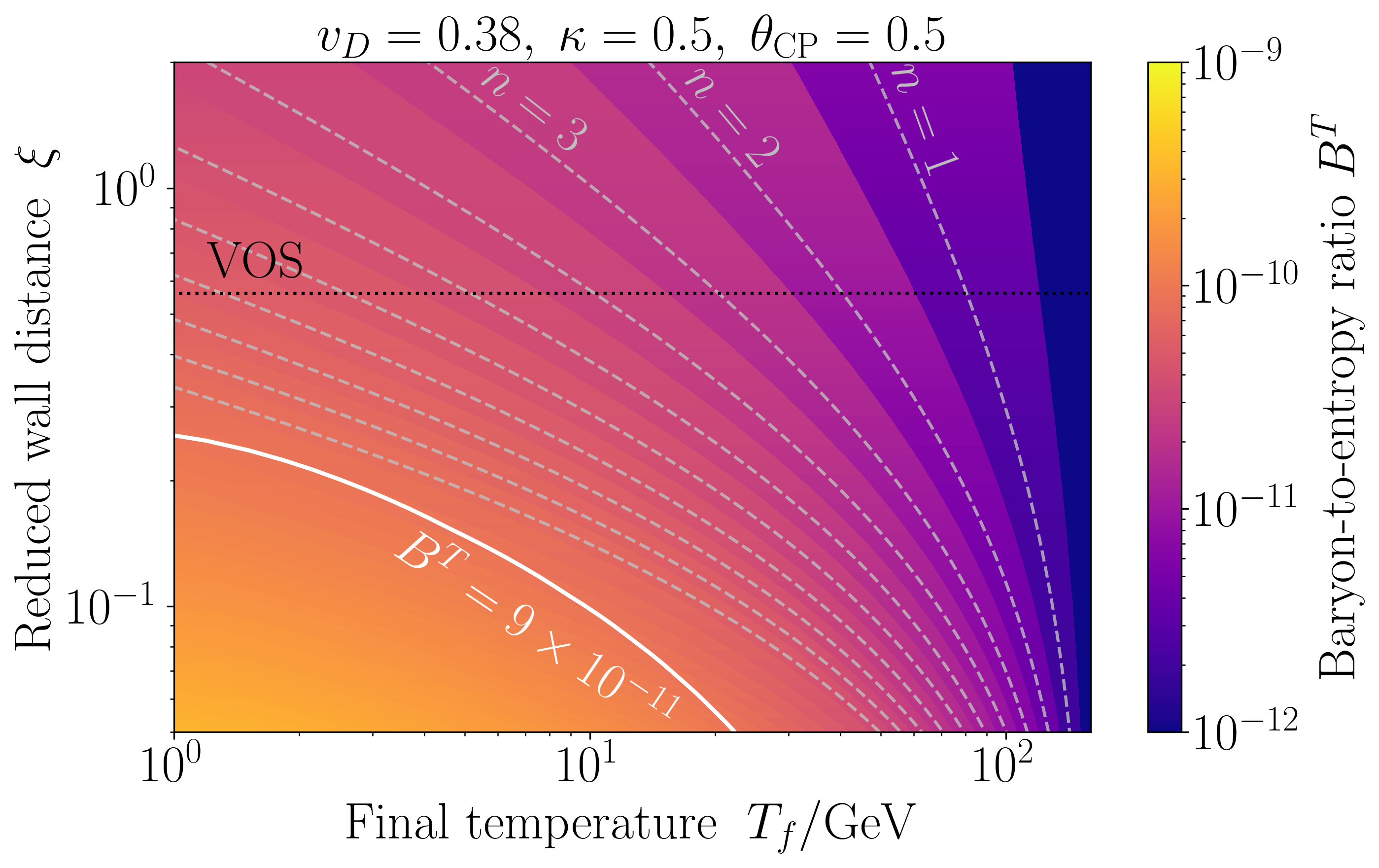}
    \includegraphics[width=0.49\linewidth]{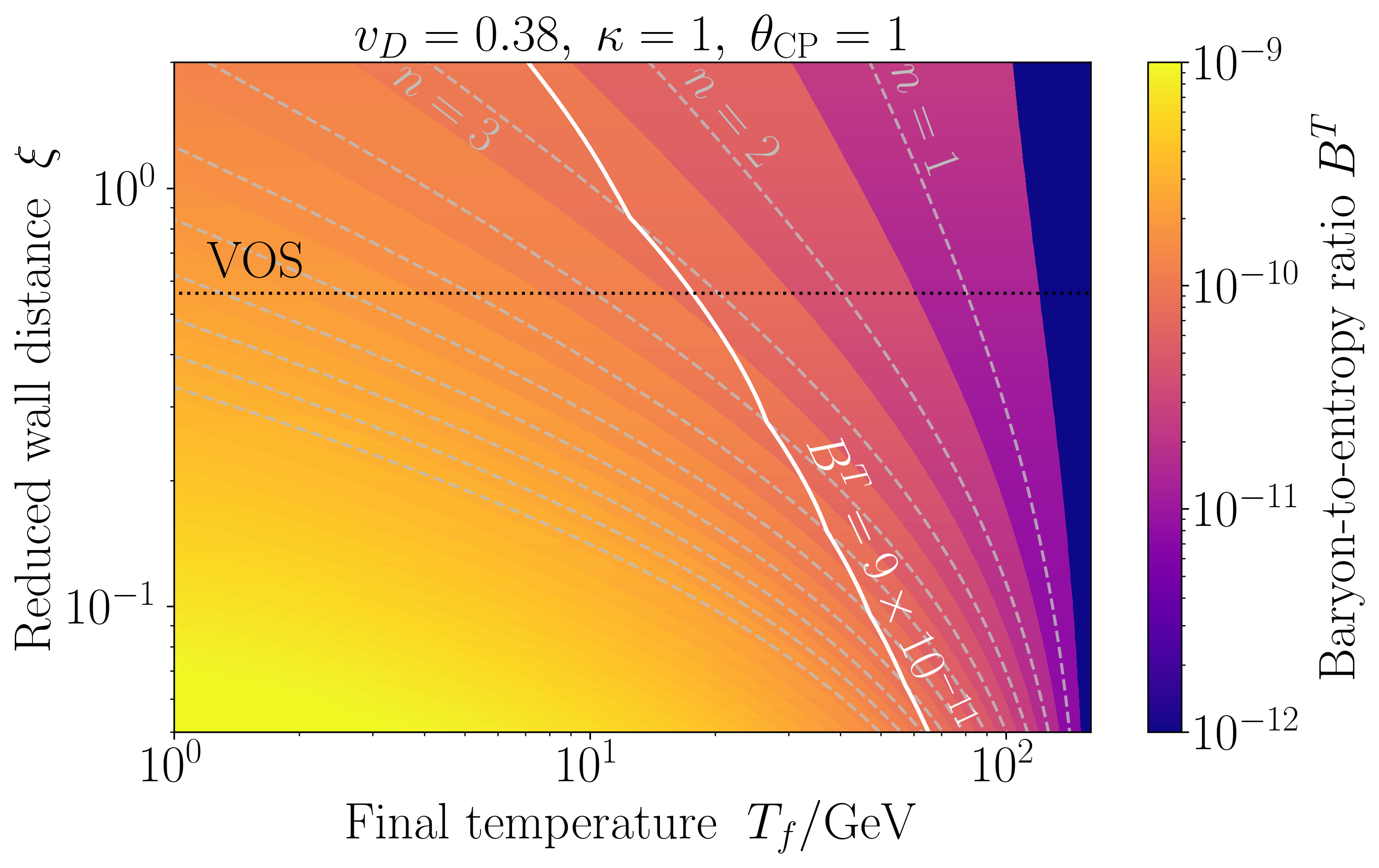}
    \includegraphics[width=0.49\linewidth]{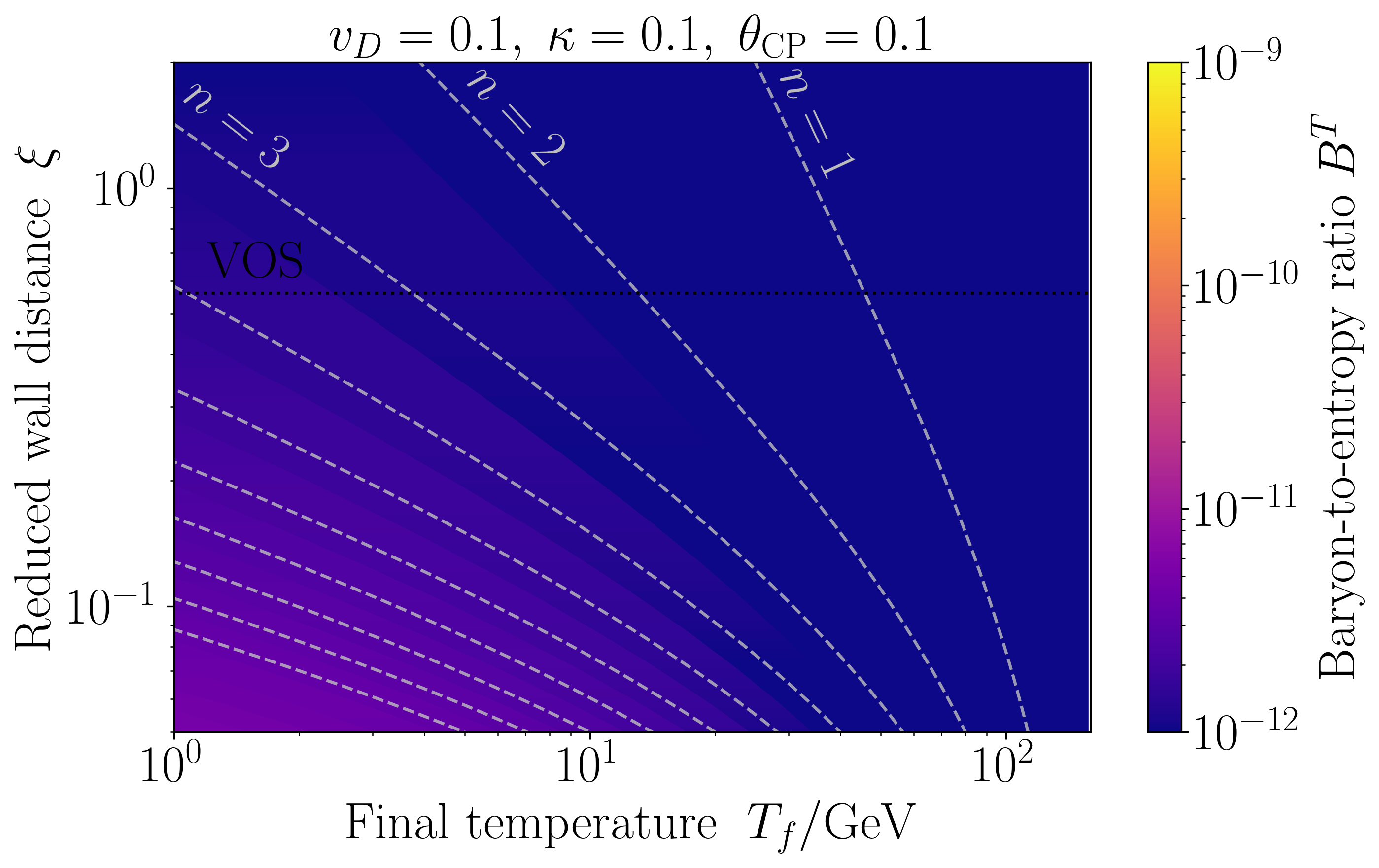}
    \includegraphics[width=0.49\linewidth]{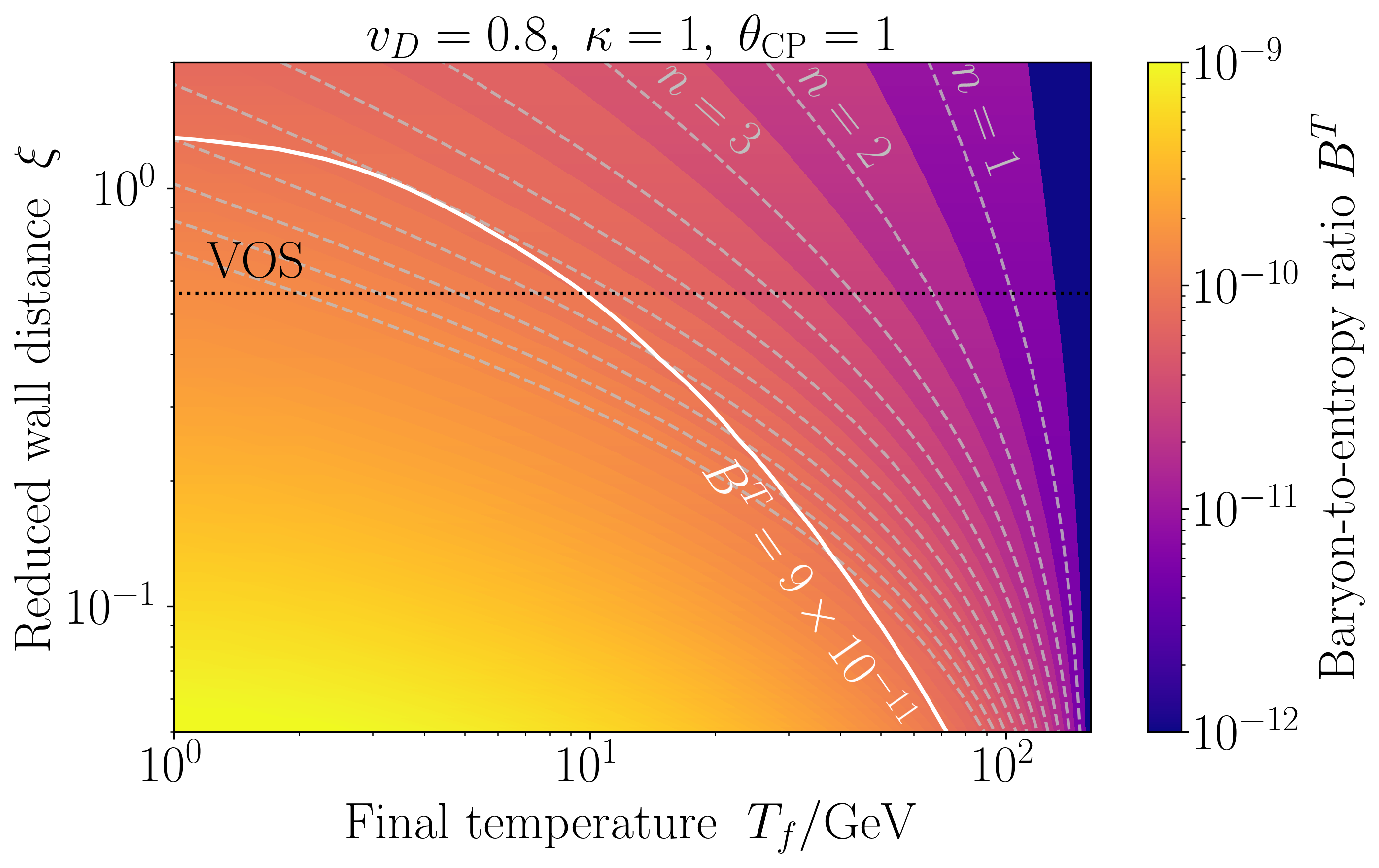}
    \caption{Plots showing the final baryon-to-entropy ratio obtained from \eqref{nonlocalBG1} with numerically calculated $\frac{n_b^0}{s}$ (cf. section \ref{section3}) as a function of the reduced average domain wall distance $\xi$ (cf. equation \eqref{eq:averageDWdistance}) and final temperature $T_f$ at which the embedded walls dissolve such that baryogenesis terminates for four different values of $(v_D, \kappa, \theta_{\rm CP})$. The wall velocity $v_D \simeq 0.38$ corresponds to the prediction of the VOS model. The black dashed line indicates the value $\xi \simeq 0.56$ predicted by the VOS model, and the solid white line shows the observed baryon-to-entropy ratio $B^T\simeq 9\times 10^{-11}$. The light-grey dashed lines show how often the DWs have swept over each point in space after $T_{\rm EW}$.}
    \label{fig:Parameter_Plot}
\end{figure*}

Equations \eqref{localBG1} and \eqref{nonlocalBG1} can be solved by 
\begin{align}
    B_0^n &= {n}_b^0\left(1 - \exp\left(-\frac{v_D l_D}{D}\right)\right) \sum_{k=1}^n \exp\left(-(n-k)\frac{\overline{\Gamma}_s l_D}{v_D}\right) \nonumber\\&= {n}_b^0\left(1 - \exp\left(-\frac{v_D l_D}{D}\right)\right)\frac{1 - \exp\left(- n\frac{\overline{\Gamma}_s l_D}{v_D}\right) }{1 - \exp\left(-\frac{\overline{\Gamma}_s l_D}{v_D}\right) }
\end{align}
in the nonlocal case, and in the local case by
\begin{align}
    B_0^n = \tilde{n}_b^0 \left(1-\exp\left(-n\frac{\overline{\Gamma}_sl_D}{v_D}\right)\right) \, ,
\end{align}
respectively.  Here, the sum runs over all wall crossings during the time when $n_b^0, \frac{v_Dl_D}{D}$, and $\frac{\bar{\Gamma}_sl_D}{v_D}$ can be taken to be approximately time-independent.

If $v_Dl_D / D \ll 1$ and $\overline{\Gamma}_sl_D / v_D \ll 1$, we can get in the non-local case an enhancement \cite{Prokopec}
\be
B_0^n \, = \, \frac{v_D^2}{\overline{\Gamma}_s D}{n}_b^0 \left(1-\exp\left(-n\frac{\overline{\Gamma}_sl_D}{v_D}\right)\right)\stackrel{n\to\infty}{\longrightarrow} \frac{v_D^2}{\overline{\Gamma}_s D}{n}_b^0 \, . 
\ee
In our case $v_D^2 / (\overline{\Gamma}_s D) \sim 10^2$. This enhancement is absent in the local case.  Note that the enhancement factor we have obtained using this consideration is comparable to the one we obtained at the beginning of this section using the more naive approach.

To make the enhancement efficient in the non-local case, we require $n \geq \, v_D / (\overline{\Gamma}_s l_D)$. Let us now express this requirement in terms of the ratio between the temperature $T_f$ when the domain wall network decays and the temperature $T_i = T_{\rm EW}$ when electroweak baryogenesis can begin. We need to estimate the number $n$ of times a given point ${\bf{x}}$ in space is crossed by a domain wall. We estimate this number using the scaling solution of the domain wall network.   
 
Taking the Hubble length to change discretely after each Hubble-time step, the time between passes over each point in space is $\tau_n = \frac{L_D(t_n)}{v_D}$ with $L_D$ the average distance between domain walls. Since $L_D(t)\propto \frac{1}{H(t)} = 2t$, which holds under the assumption that we are in the scaling regime and deep in the radiation-dominated era, we can write 
\be L_D(t) = 2\xi t \label{eq:averageDWdistance} \ee 
and hence $\tau_n = \frac{2\xi t_n}{v_D}$. The time after the $n$-th passage over each point in space is 
\be
t_n = t_{n-1} + \tau_{n-1} = t_{n-1}\left(1+\frac{2\xi}{v_D}\right) \label{eq:tn}
\ee
and thus 
\be
t_n = \left(1+\frac{2\xi}{v_D}\right)^nt_0 \, . 
\ee
Using that during radiation domination 
\be
\frac{1}{2t}= H(t) = \sqrt{\frac{8\pi^3 G}{90}g^*} T^2 
\ee
holds, this translates to under the assumption of approximately constant degrees of freedom to 
\begin{align}
T_n = \frac{T_0}{\left(1+\frac{2\xi}{v_D}\right)^{n/2}} \label{Temperaturenrelation}
\end{align}
and, therefore, the enhancement in the case of non-local baryogenesis becomes efficient when
\begin{align}
    \frac{T_i}{T_f} \geq \left(1+\frac{2\xi}{v_D}\right)^{\frac{v_D}{2\overline{\Gamma}_s l_D}}.
\end{align}
Finally, let us mention that in order to apply the above treatment to much lower temperatures than $m_H$, for more precise values of $B^T$, numerical evaluation of the integrals \eqref{eq:generalexpressionlefthandedflux} and \eqref{particlevelocities} is necessary and shows that our mechanism can for reasonable parts of the parameter space comfortably explain the observed baryon-to-entropy ratio. Results of this numerical evaluation are shown in Fig. \ref{fig:Parameter_Plot}. The four plots depict the total baryon-to-entropy ratio $B^T$ after the baryon production has ceased as a function of the reduced average domain wall distance $\xi=\frac{L_D(t)}{2t}$ and the temperature $T_f$ at which the embedded DWs decay and baryogenesis stops. In each plot, we consider a different set of parameters $v_D, \, \kappa,$ and $\theta_{\rm CP}$. As a common feature of all four plots, lower $\xi$ and lower $T_f$ lead to a higher value of $B^T$. This is because both lead to an increase in the number of times each point in space is washed over by a domain wall before the embedded walls decay. The parameters $\kappa$ and $\theta_{\rm CP}$ enter almost exclusively as a common prefactor $\kappa \theta_{\rm CP}$ in $B^T$. The role of $v_D$ is more complicated. For very low values, the domain walls wash over each point in space only a few times before they decay, leading to a small $B^T$. On the other hand, a too large value of $v_D$ can reduce the baryon number produced in a single passage of a wall. The solid white line in the plots indicates the observed baryon-to-entropy ratio. As can be seen, for a large part of the displayed parameter space, the presented mechanism leads to a sufficient baryon production to explain observations. It should be emphasized again that the parameter dependencies and numbers shown in Fig. \ref{fig:Parameter_Plot} are based on order-of-magnitude estimations. 

 \section{A Model with Embedded Walls} 
\label{section5} 

While the Standard Electroweak Model has embedded strings,  it is necessary to go beyond the Standard Model to obtain embedded walls.  Here, we will propose an extension of the Standard Model, which admits embedded walls within which the electroweak symmetry is restored.

We are looking for an extension of the Standard Model gauge symmetry group such that, when the enhanced symmetry group $G$ breaks to the gauge group of the Standard Model, embedded defects arise.  Thus,  we consider the symmetry breaking pattern \footnote{$SU(3)_C$ is the gauge group of the strong interactions while $SU(2)_L \times U(1)_Y$ is the electroweak symmetry group which is broken by the standard Higgs field to the $U(1)_{\rm EM}$ symmetry of electromagnetism.}
\be
G\stackrel{\Phi}{\longrightarrow} SU(3)_C \times SU(2)_L \times U(1)_Y \stackrel{H}{\longrightarrow} SU(3)_C\times U(1)_{\rm EM} \label{eq:BreakingChain}
\ee
where $\Phi$ is a new Higgs field that acquires a non-vanishing vacuum expectation value (VEV) in the first phase transition, and $H$ is the electroweak Higgs, which may or may not be embedded in $\Phi$. The symmetry breaking should occur in such a way that the vacuum manifold $\mathcal{M}$ is connected, i.e., $\pi_0\left(\mathcal{M}\right)\cong \{1\}$, and hence admits no topologically stabilized domain walls. It should, however, allow for embedded domain wall solutions, which can be stabilized via interactions with the thermal plasma. We might, e.g., have some Higgs field $\Phi = (\phi_1, \dots, \phi_{n+1})$, such that its vacuum is described by 
\be
\mathcal{M} \, = \, \left\{\Phi\in \mathbb{R}^{n+1}|\sum_{i=1}^{n+1} \phi_i^2 = v_\Phi^2\right\}\, \cong \, S^n \, . 
\ee
It is important to note that the $\phi_i$ are real-valued. If the fields $\phi_1, \dots, \phi_n$ are charged under $U(1)_{\rm EM}$ or $SU(3)_C$ while $\phi_{n+1}$ is uncharged, interactions of the fields with the thermal plasma in the early universe can lift the potential in the direction of the charged fields, effectively reducing the vacuum to 
\be
\mathcal{M}_{\rm eff} \, = \, \left\{\Phi\in \mathbb{R}^{n+1}|\phi_{n+1}=\pm v_\Phi\right\} \, \cong \,  S^0
\ee
which is disconnected. These interactions stabilize embedded domain wall solutions in which all of the $\phi$ fields except for $\phi_{n + 1}$ vanish, and $\phi_{n+1}$ transits from $-v_{\Phi}$ to $+v_{\Phi}$.  Notice that this is only possible if $n$ is even.  

As a simple realization of this scenario, we propose to study a model in which $U(1)_Y$ is embedded in a $SU(2)_\sfA\times U(1)_\sfB$ symmetry group
\be
G \, = \, SU(3)_C\times SU(2)_L\times SU(2)_\sfA\times U(1)_\sfB. \label{eq:FullGroup}
\ee
Since $SU(3)_C$ remains unbroken in the chain \eqref{eq:BreakingChain}, we do not have to consider it in detail and will ignore it for the following discussion. The vacuum manifold, after the entire symmetry breaking, is connected, admitting, therefore, no topologically stable domain walls. 

Symmetry groups of a similar form as $G$ are, for example, known from left-right symmetric models (see, e.g., \cite{Mohapatra}).

In the low-energy limit (at the electroweak scale), we need to recover the Standard Model Lagrangian, in particular for the gauge and Higgs fields
\begin{align}
    \mathcal{L}_{\rm SM} \, \supset \, \mathcal{L}_{\rm kin}^{0} + \mathcal{L}_H^0
\end{align}
where 
\be
\mathcal{L}_{\rm kin}^0 \, = \,  -\frac{1}{2}\text{tr}\left(W_{\mu\nu}W^{\mu\nu}\right) -\frac{1}{4}B^{\mu\nu}B_{\mu\nu}
\ee
describes the kinetic terms of the electroweak gauge fields, namely the $W$-bosons $W^\mu$ associated with $SU(2)_L$-symmetry and with field strength $W^{\mu\nu}$, and the $B$-boson $B^\mu$ associated with $U(1)_Y$-symmetry and field strength $B^{\mu\nu}$. The electroweak Higgs $H$ is a complex $SU(2)_L$ doublet with weak hypercharge $Y(H)=1/2$
\be
\mathcal{L}_H^0  \, = \, \left(\mathcal{D}_\mu^0 H\right)^\dagger \left(\mathcal{D}^\mu_0 H\right) - V_H(H) 
\ee
with 
\be
V_H(H) \, = \, \lambda_H\left(H^\dagger H - v_H^2\right)^2 \label{EWHiggs}
\ee
and
\be
    \mathcal{D}_\mu^0 H \, = \,  \left(\partial_\mu - igW_\mu^a \tau^a - \frac{i}{2}g'B_\mu\right) H \, ,
\ee
where $H = \left(H^+, H^0\right)$ and the generators being the re-scaled Pauli matrices $\tau^a = \sigma^a / 2$. The label $0$ on the covariant derivative is to distinguish the covariant derivative in the SM from the covariant derivative of the full symmetry group.

Starting from a phase with the full internal symmetry, the breaking to the Standard Model must involve $SU(2)_\sfA\times U(1)_\sfB \longrightarrow U(1)_Y$. To obtain embedded domain walls, we realize this symmetry breaking in two steps
\begin{align}
    SU(2)_\sfA\times U(1)_\sfB \stackrel{\Phi}{\longrightarrow} U(1)_\sfA\times U(1)_\sfB \stackrel{\Psi}{\longrightarrow} U(1)_Y.
\end{align}
Here, $\Phi$ is taken to be in the $(1,1,3)_0$ representation of $G$ and $\Psi$ is in the $(1,1,2)_{\frac{1}{2}}$ representation. With the Pauli matrices $\sigma^A$, we can denote the $SU(2)_\sfA$-generators $t^A = \sigma^A/2$ which then have normalization $\tr\left(t^A t^B\right) = \frac{1}{2}\delta^{AB}$. Hence, we can write $\Phi$ in terms of three real fields $\phi^A$ ($A = 1,2, 3$) as
\be
    \Phi \, = \, \phi^A t^A = \frac{1}{2}\begin{pmatrix} \phi^3 && \phi^1 - i \phi^2 \\ \phi^1 + i \phi^2 && -\phi^3\end{pmatrix} 
\ee
and $\Psi$ as a complex doublet
\be
\Psi \, = \, \begin{pmatrix}
        \psi_1 \\ \psi_2
    \end{pmatrix}. 
\ee
To fix conventions, let us also introduce the gauge connections $R_\mu = R_\mu^A t^A$ for the $SU(2)_\sfA$-group and $S_\mu$ for the $U(1)_\sfB$-group. The covariant derivatives of the scalars are then given by
\begin{align}
    \mathcal{D}_\mu \Phi = \p_\mu \Phi - ig_\sfA [R_\mu, \Phi]\, , \\
    \mathcal{D}_\mu \Psi = \p_\mu \Psi - ig_\sfA R_\mu \Psi - \frac{i}{2}g_\sfB S_\mu \Psi \,.
\end{align}
Under gauge transformations 
\be
U_\sfA(x) \, = \, \exp\left(i\alpha(x)\right) \,\,\, {\rm{with}} \,\,\, \alpha(x) \, = \, \alpha^A(x) t^A
\ee
and 
\be
U_\sfB(x) \, = \, \exp\left(\frac{i}{2}\beta(x)\right),
\ee
the gauge and scalar fields transform as
\begin{align}
  R_\mu &\longmapsto U_\sfA R_\mu U^{-1}_\sfA + \frac{i}{g_\sfA}U_\sfA\p_\mu U_\sfA^{-1},\\
  S_\mu &\longmapsto S_\mu  + \frac{1}{g_\sfB}\p_\mu \beta,\\
    \Psi&\longmapsto U_\sfA U_\sfB \Psi, \\
    \Phi &\longmapsto U_\sfA \Phi U_\sfA^{-1}.
\end{align}

Let us now realize the first symmetry stage of symmetry breaking with a potential 
\ba
    V_1(\Phi) \, &=& \, \lambda_\Phi \left(\tr\left( \Phi^2\right)-\frac{1}{2} v_\Phi^2\right)^2 \nonumber \\
    &=& \frac{\lambda_\Phi}{4} \left(\sum_{A=1}^3 \left(\phi^A\right)^2 - v_\Phi^2\right)^2. \label{eq:VPhi}
\ea
The minimum of the potential is clearly at
\be
\sum_{A=1}^3 \left(\phi^A\right)^2 = v_\Phi^2
\ee
and the corresponding vacuum manifold  has, therefore, the topology 
\begin{align}
\mathcal{M}_\Phi \cong S^2. \label{eq:PhiVacuum}
\end{align}
Locally, we can choose our $3$-axis in group space to be parallel to the vacuum expectation value of $\Phi$ such that $\Braket{\Phi} = v_\Phi t^3 $. The remaining unbroken generator is then $t^3$, corresponding to an unbroken $U(1)_\sfA$-subgroup. Note that the remaining symmetry group acts now on $\Psi$ as
\begin{align}
    \begin{pmatrix}
        \psi_1(x)\\ \psi_2(x)
    \end{pmatrix} \longrightarrow \begin{pmatrix}
        e^{\frac{i}{2} \left(\alpha^3(x)+ \beta(x)\right)} \psi_1(x) \\
        e^{\frac{i}{2}\left(-\alpha^3(x)+\beta(x)\right)}\psi_2(x) \label{eq:U1U1}
    \end{pmatrix}.
\end{align} 

For the second stage of symmetry breaking, we can again use a quartic potential 
\begin{align}
    V_2(\Psi) \, = \,  \lambda_\Psi \left(\Psi^\dagger \Psi - \eta_\Psi^2\right)^2.
\end{align}
In this way, both $\psi_1$ and $\psi_2$ will acquire non-vanishing VEVs which  would break the remaining $U(1)_\sfA\times U(1)_\sfB$-symmetry completely. 

To make sure that only one of the two acquires a non-vanishing VEV, we have to lift the degeneracy between $\psi_{1}$ and $\psi_2$ by breaking $SU(2)_\sfA$ in the $\Psi$-potential, i.e., we need to introduce interactions between $\Phi$ and $\Psi$. This can be achieved via the gauge-invariant term
\begin{align}
    V_3(\Phi, \Psi)\, = \, M \Psi^\dagger \Phi \Psi \, \ceq \, \frac{M v_\Phi}{2}\left(\left|\psi_1\right|^2 -\left|\psi_2\right|^2\right),
\end{align}
where $\ceq$ means that the expression on the left-hand side becomes effectively the expression on the right-hand side below the symmetry-breaking scale $\sim v_\Phi$. Later on, we will use $\cceq$ to mean a similar thing but with respect to the second symmetry breaking at scale $v_\Psi$ and $\stackrel{\circ\circ\circ}{=}$ for the electroweak symmetry breaking. 

The potential of $\Psi$ becomes then effectively 
\ba
    V_2(\Psi) + V_3(\Phi,  \Psi) \, &\ceq& \, \lambda_\Psi |\psi_1|^4 + \lambda_\Psi |\psi_2|^4   + 2\lambda_\Psi |\psi_1|^2|\psi_2|^2 \nonumber \\
    & & + \mu_1 |\psi_1|^2 - \mu_2 |\psi_2|^2 + \lambda_\Psi v_\Psi^4, 
\ea
where we defined 
\ba
\mu_1 \, &=& \, \frac{Mv_\Phi}{2} - 2\lambda_\Psi \eta_\Psi^2 \,\,\, {\rm{and}} \\
\mu_2 \, &=& \,  \frac{Mv_\Phi}{2}+2\lambda_\Psi \eta_\Psi^2 \, . \nonumber
\ea
The part of parameter space that is interesting for us (and which is the natural one) is $\mu_{1,2}>0$. The minimum of the potential that needs to be considered below the first symmetry-breaking scale lies at
\ba
    |\psi_1|^2 \, &=& \, 0  \,\,\, {\rm{and}} \\
   |\psi_2|^2 \, &=& \,  \frac{\mu_2}{2\lambda_\Psi}\equiv v_\Psi^2.  \nonumber
\ea
The vacuum associated with this is
\begin{align}
    \mathcal{M}_\Psi \, \cong \, S^1
\end{align}
and we can choose the field space direction again such that
\begin{align}
 \Braket{\Psi} =   \begin{pmatrix}
        0 \\ v_\Psi
    \end{pmatrix}.
\end{align}
Transformations affecting the $\psi_2$-component are therefore no longer symmetries.

Looking at \eqref{eq:U1U1}, we can however see that transformations with $\alpha^3(x)=\beta(x)$ leave $\psi_2$ untouched and the remaining symmetry is a $U(1)$ symmetry that acts on $\psi_1(x)$ as
\begin{align}
    \psi_1(x) \longrightarrow e^{i\beta(x)}\psi_1(x).
\end{align}
We can associate this with weak hypercharge $U(1)_Y$-symmetry and the corresponding charge operator is
\begin{align}
   Q_Y \, =\,  t^3 + Q_\sfB \label{eq:YchargeOperator}
\end{align}
where $Q_\sfB$ is the corresponding $U(1)_\sfB$-charge and $t^3$ the unbroken $SU(2)_\sfA$-generator. 

We are now able to determine the electric charges of $\Phi$ and $\Psi$. Since both are in the trivial representation of $SU(2)_L$, their electric charge $Q$ is equivalent to their weak hypercharge. For $\Psi$, we can directly see that
\begin{align}
    Q\left(\psi_1\right) = 1 && \text{and} && Q\left(\psi_2\right) = 0.
\end{align}
The latter is important since $\psi_2$ acquires a non-vanishing VEV. It would break $U(1)_{\rm EM}$ if it had a non-vanishing electric charge. For $\Phi$, we have to determine the eigenstates corresponding to the unbroken generator $t^3$. We can write
\begin{align}
    t^\pm = \frac{1}{2}\left(t^1 \pm it^2\right)
\end{align}
which satisfy $[t^3, t^\pm]=\pm t^\pm$. Correspondingly, we find 
(with the decomposition $\Phi = \phi^+ t^+ + \phi^- t^- + \phi^0 t^3$) that the transformation behaviour under the unbroken $U(1)_\sfA$ is
\ba
    \phi^+(x) &\longrightarrow& e^{i\alpha^3(x)}\phi^+(x),  \nonumber \\
    \phi^-(x) &\longrightarrow& e^{-i\alpha^3(x)}\phi^-(x),  \\
     \phi^0(x) &\longrightarrow & \phi^0(x) \, , \nonumber 
\ea
and, therefore, the electric charges are
\ba
    Q\left(\phi^+\right) &=& 1,  \nonumber \\
     Q\left(\phi^-\right) &=& -1,  \\
     Q\left(\phi^0\right) &=&  0. \nonumber
\ea

As we will discuss in more detail below, interactions between $\Phi$ and thermal electromagnetic radiation will effectively change the potential \eqref{eq:VPhi} by lifting it in the $\phi^1$ and $\phi^2$ directions such that the remaining vacuum has effectively the topology
\begin{align}
    \mathcal{M}_\Phi^{\rm eff} \cong S^0
\end{align}
and thus stabilizes (embedded) domain wall solutions.

It is important to specify how the Standard Model (SM) particles can be included in our  model. Since the $SU(3)_C\times SU(2)_L$-part of the SM is not changed by the larger symmetry group, the representations of all SM particles under these subgroups stay the same. We can furthermore make all SM particles singlets under $SU(2)_\sfA$ and assign them $U(1)_\sfB$-charges which are the same as their $U(1)_Y$-charges according to \eqref{eq:YchargeOperator}. Therefore, the particle content of the new theory stays the same as in the SM except for the introduction of $\Phi, \Psi$, and four gauge bosons of the $SU(2)_\sfA\times U(1)_\sfB$-symmetry, of which one field becomes the electroweak $B$-boson of the $U(1)_Y$-symmetry, though.

The electroweak symmetry breaking $SU(2)_L\times U(1)_Y\stackrel{H}{\longrightarrow}U(1)_{\rm EM}$ is obtained with the usual electroweak Higgs, an $SU(2)_L$ doublet with weak hypercharge $\frac{1}{2}$.  So far, we have obtained a theory with embedded domain walls,  but we must ensure that the electroweak symmetry is restored inside of them.  This is possible by introducing a coupling between $H$ and $\Phi$ via
\begin{align}
    V_4(\Phi, H) \, = \, \sigma H^\dagger H \left(\frac{v_\Phi^2}{2} - \tr\left(\Phi^2\right)\right) \, , \label{eq:PhiEWHiggscoupling}
\end{align}
where $\sigma$ is a positive constant. Outside the domain wall, where $SU(2)_\sfA$ is broken, $V_4=0$ and the usual electroweak theory remains unaffected, while inside the domain wall, where the symmetry is unbroken, one obtains 
\begin{align}
V_4 \, = \, \frac{1}{2}\sigma v^2_\Phi H^\dagger H \,
\end{align}
which contributes to the electroweak Higgs potential (cf. \eqref{EWHiggs})
\begin{align}
    V_4(\Phi, H) + V_H(H) \ceq\,\,\,\,\,\,\,\,\,& \\ \nonumber \ceq \left(\frac{1}{2}\sigma v^2_\Phi - 2\lambda_H v_H^2\right)& H^\dagger H +\lambda_H (H^\dagger H)^2 + \lambda_H v_H^4. 
\end{align}
Therefore, the minimum of the potential is at $H = 0$ for $v_\Phi^2 \geq \frac{4\lambda_H}{\sigma} v_H^2$, and the electroweak symmetry is unbroken inside the domain wall.

Let us now go through the different symmetry breaking steps, the corresponding Higgs mechanism and determine the corrections to the effective potential coming from plasma interactions. We start with the first symmetry breaking. The kinetic term for the first Higgs field reads
\begin{widetext}
\begin{align}
    \tr\left[\left(\mathcal{D}_\mu \Phi\right) \left(\mathcal{D}^\mu \Phi\right)\right] &= \tr\left(\p_\mu \Phi \p^\mu \Phi\right) - 2ig_\sfA \tr\left(\p_\mu \Phi [R^\mu, \Phi]\right) - 2g_\sfA^2\tr\left(R_\mu \Phi R^\mu \Phi\right) + 2g_\sfA^2 \tr\left(R_\mu R^\mu \Phi^2\right)\stackrel{\circ}{\supset} \nonumber\\ &\stackrel{\circ}{\supset} \frac{g_\sfA^2 v_\Phi^2}{2}\left(R_\mu^1 R_1^\mu + R_\mu^2 R^\mu_2\right).   \label{Higgsmech1}
\end{align}
\end{widetext}
Correspondingly, the two gauge bosons $R_\mu^{1,2}$ gain masses while $R_\mu^3$ remains massless
\begin{align}
    m_R^{1,2} \ceq g_\sfA v_\Phi, && m_R^{3} \, = \, 0.
\end{align}

Let us now consider the universe after the first symmetry breaking. Since the photon and hence the $R^3$ gauge field are in thermal equilibrium, we can apply thermal averages and approximate \cite{Nagasawa}
\begin{align}
\Braket{R_\mu^3}_T \simeq 0 && \text{and} && \Braket{R_\mu^3 R^\mu_3}_T \simeq -\kappa_R T^2
\end{align}
with an order one prefactor $\kappa_R$. The quadratic terms in the first line of \eqref{Higgsmech1} give then rise to a contribution to the effective Lagrangian of the form
\be 
-\frac{g_\sfA^2 \kappa_R}{2} T^2\left(\left(\phi^1\right)^2+\left(\phi^2\right)^2\right), \label{eq:effectivemass}
\ee
and the fields $\phi^{1,2}$ acquire, therefore, the masses
\begin{align}
    m_{\phi}^{1,2} \, \stackrel{\circ}{\simeq} \, \sqrt{\kappa_R}g_\sfA T.
\end{align}
We can now see explicitly that the term \eqref{eq:effectivemass} yields an effective contribution to the potential \eqref{eq:VPhi} and lifts it in the charged field directions. Thus, the remaining vacuum becomes $S^0$ instead of $S^2$ and embedded domain walls are stabilized. 

As usual, the new physical field $\varphi \equiv \phi^3-v_\Phi$ obtains a mass
\begin{align}
    m_\varphi \, = \, \sqrt{2\lambda_\Phi} v_\Phi.
\end{align}

Let us now come to the second symmetry-breaking. Introducing 
\begin{align}
    \begin{pmatrix} P_\mu \\ B_\mu \end{pmatrix} = \frac{1}{\sqrt{g_\sfA^2+g_\sfB^2}} \begin{pmatrix}
        g_\sfA && - g_\sfB\\ g_\sfB && g_\sfA
    \end{pmatrix}\begin{pmatrix} R_\mu^3 \\ S_\mu \end{pmatrix}, 
\end{align}
we can write for the kinetic term of the second Higgs field
\begin{widetext}
\begin{align}
    \left(\mathcal{D}_\mu \Psi\right)^\dagger \left(\mathcal{D}^\mu \Psi\right) \stackrel{\circ\circ}{\supset} \frac{g_\sfA^2v_\Psi^2}{4} \left(R_\mu^1R^\mu_1+R_\mu^2R^\mu_2\right) + \frac{\left(g_\sfA^2+g_\sfB^2\right)v_\Psi^2}{4} P_\mu P^\mu.
\end{align}
\end{widetext}
We have then the following masses for our gauge fields
\begin{align}
    m_R^{1,2}\cceq g_\sfA\sqrt{v_\Phi^2 + \frac{v_\Psi^2}{2}}, && m_P = \sqrt{\frac{g_\sfA^2+g_\sfB^2}{2}}v_\Psi, && m_B = 0.  
\end{align}
Since we consider the unbroken $U(1)$-symmetry as corresponding to the weak hypercharge, $B_\mu$ is the $B$-boson of electroweak interactions. As one can easily check  by considering how a field in an arbitrary representations of $SU(2)_{\sfA}\times U(1)_{\sfB}$ with $SU(2)_\sfA$-generators $T^A$ transforms, fixing 
\begin{align}
g' = \frac{g_\sfA g_\sfB}{\sqrt{g_\sfA^2+g_\sfB^2}}
\end{align}
to be the $B$ gauge coupling, the weak hypercharge operator reads 
\begin{align}
    Q_Y = T^3 + Q_\sfB.
\end{align}
We can apply again thermal averages after the second symmetry breaking, in which case we have now $\Braket{B_\mu B^\mu}_T \simeq -\kappa_B T^2$ which leads via \eqref{Higgsmech1} to a term $-\frac{g'^2\kappa_B}{2} T^2 \left(\left(\phi^1\right)^2+\left(\phi^2\right)^2\right)$ such that $\phi^{1,2}$ have after the second symmetry breaking the thermal masses 
\begin{align}
    m_\phi^{1,2} \stackrel{\circ\circ}{\simeq} \sqrt{\kappa_B}g' T.
\end{align}
Similarly, the thermal average leads to a term $-g'^2 \kappa_B T^2 |\psi_1|^2$ such that also $\psi_1$ acquires the same thermal mass
\begin{align}
    m_\psi^1 \stackrel{\circ\circ}{\simeq} \sqrt{\kappa_B }g' T. 
\end{align}

Let us finally consider electroweak symmetry-breaking $SU(2)_L\times U(1)_Y \to U(1)_{\rm EM}$. The Higgs field $H$ belongs to the $(1,2,1)_{\frac{1}{2}}$ representation of the full group \eqref{eq:FullGroup} and transforms correspondingly as $H(x)\to e^{i\gamma^a(x) \tau^a}e^{\frac{i}{2}\beta(x)}H(x)$. From the potential \eqref{EWHiggs}, we can read off that, outside the domain walls, the minimum of the potential is at $H^\dagger H = v_H^2$. We can choose the VEV correspondingly in the usual way
\begin{align}
   \Braket{H} = \begin{pmatrix}
       0 \\ v_H
   \end{pmatrix}. 
\end{align}
This VEV is invariant under $SU(2)_L\times U(1)_Y$-transformations with $\gamma^{1,2}(x)=0$ and $\gamma^3(x)=\beta(x)$ corresponding to a $U(1)$ subgroup which we interpret as the gauge group of electrodynamics.\\ Let us now look at the mass generation. From the kinetic term of the electroweak Higgs, we can find that the linear combination 
\begin{align}
    A_\mu = \frac{g' W_\mu^3 + g B_\mu}{\sqrt{g^2+g'^2}}
\end{align}
is the only gauge field which stays massless and can correspondingly be interpreted as the gauge boson of the unbroken $U(1)_{\rm EM}$, i.e., the photon. Considering how a field in an arbitrary representation of $SU(2)_L\times SU(2)_{\sfA}\times U(1)_{\sfB}$ with $SU(2)_L$ generator $T_L^3$ and $SU(2)_{\sfA}$ generator $T^3$ transforms, 
one can check that the electromagnetic coupling constant and charge operator can be defined consistently via
\begin{align}
    e &= \frac{gg'}{\sqrt{g^2+g'^2}}, \\ Q &= T_L^3 + Q_Y = T_L^3 + T^3 + Q_B. \nonumber
\end{align}
After electroweak symmetry breaking, we can apply the thermal average $\Braket{A_\mu A^\mu}_T\simeq-\kappa_A T^2$ and find $\tr\left[\left(\mathcal{D}_\mu \Phi\right)\left(\mathcal{D}^\mu \Phi\right)\right] \supset -\frac{e^2}{2}\kappa_A T^2 \left(\left(\phi^1\right)^2+\left(\phi^2\right)^2\right)$ such that
\begin{align}
m_\phi^{1,2}\stackrel{\circ\circ\circ}{\simeq} \sqrt{\kappa_A}e T.
\end{align}
As we can see, plasma interactions lift the potential \eqref{eq:VPhi} also after electroweak symmetry breaking in the $\phi_{1,2}$ directions, the effective vacuum remains disconnected and embedded domain walls are stabilized. 
Similarly, we find for the other Higgs fields the terms $-e^2 \kappa_A T^2 |\psi^1|^2$ and $-e^2 \kappa_A T^2 |H^+|^2 $ in the Lagrangian such that these fields acquire additional mass contributions as well
\begin{align}
m_\psi^1 \stackrel{\circ\circ\circ}{\simeq} \sqrt{\kappa_A} e T &&     m_H^+  \stackrel{\circ\circ\circ}{\simeq} \sqrt{\kappa_A} e T.
\end{align}

In the above model, we have not yet specified how CP symmetry is violated at the domain wall boundary. To directly apply the discussion of section \ref{section3}, CP violation should be included via a two-Higgs-doublet model (see, e.g. \cite{localBG}). The second Higgs can be coupled to the Higgs field $\Phi$, which makes up the embedded domain walls, analogous to the first Higgs doublet $H$ in equation \eqref{eq:PhiEWHiggscoupling}.

\section{Discussion}  
 \label{section6}
 
We have studied the baryon-to-entropy ratio which can be induced by a network of embedded domain walls within which the electroweak symmetry is restored.  The walls represent configurations which are out of thermal equilibrium.  Thus,  in the presence of sufficient CP violation, the Sakharov conditions for baryogenesis are satisfied.  We have shown that the measured net baryon-to-entropy ratio can be obtained.   The main reason why wall-mediated baryogenesis is more efficient than the string-mediated process is that that network of walls pass over a fraction of order one of the space-time volume while the string worldsheets only cover a small fraction.

However, for our mechanism to work, certain conditions have to be satisfied. First of all, to avoid the domain wall problem, our embedded walls must have decayed. As long as the decay temperature $T_d$ is higher than the temperature of nucleosynthesis, the domain wall problem can be avoided.  The embedded wall network will decay once the plasma effects become ineffective in lifting the vacuum manifold in the charged Higgs field direction. This process needs to be carefully studied.

Let us turn to a second requirement:
The baryon number violation is provided by the usual electroweak sphalerons. For these processes to be efficient, sphalerons must fit into the walls \footnote{This is a conservative estimate. The region of symmetry restoration around the wall might be larger than the wall itself.}.  The radius $R_{\rm sph}$ at a temperature $T$ is
 \be
 R_{\rm sph} \, \sim \, \left(g^2 T\right)^{-1} \, ,
 \ee
 where $g$ is the gauge coupling constant.  The wall thickness $R_w$, on the other hand, is
 \be
 R_w \, \sim \, \lambda^{-1/2} \eta^{-1} \, ,
 \ee
 where $\lambda$ is the self-coupling constant of the Higgs field which yields the embedded walls, and $\eta$ is the corresponding symmetry breaking scale.  A requirement for our mechanism to be effective is $R_{\rm sph} < R_w$ which requires
 \be \label{bound1}
 g^{-2} T^{-1} \, < \, \lambda^{-1/2} \eta^{-1} \, 
 \ee
 evaluated at the temperature $T_{\rm EW}$ of electroweak symmetry breaking. The value of $\eta$ has to be consistent with the symmetry breaking temperature $T_c$ for the embedded wall formation being higher than $T_{\rm EW}$ and can be estimated as
 \be
 T_c \, \sim {\tilde{g}}^{-1} \lambda^{1/2} \eta \, ,
 \ee
 where $\tilde{g}$ is the coupling constant between the embedded wall field and the standard model fields. Hence, the bound (\ref{bound1}) becomes
 \be
 g^{-2} T^{-1} \, < {\tilde{g}}^{-1} T_c^{-1}
 \ee
 which can be realized for $T_c > T = T_{\rm EW}$ if ${\tilde{g}}$ is sufficiently small.   This is a second reason why embedded wall-mediated electroweak baryogenesis can be more efficient than the string-mediated process where \cite{Moore} typically a spherical sphaleron does not fit into the string core.
  
 We have discussed electroweak baryogenesis from embedded walls in which the electroweak symmetry is unbroken. The same mechanism also applies to other types of domain wall scenarios in which the walls decay at some late time (between the time of electroweak symmetry breaking and nucleosynthesis).  A possible realization is a scenario in which domain walls form at some early times in a phase transition with a disconnected vacuum manifold, but this vacuum manifold gets lifted in a later phase transition, leaving a unique vacuum behind. (see, e.g., \cite{Jim} for a generic model). Another  scenario in which our mechanism could apply is in a setup with biased \cite{biased} or metastable \cite{metastable} domain walls.
 
\begin{acknowledgements}

The research of R.B. at McGill is supported in part by funds from NSERC and from the Canada Research Chair program. The work of T.S. is supported by Deutsche Forschungsgemeinschaft (DFG) through the Research Training Group (Graduiertenkolleg) 2149: Strong and Weak Interactions\,---\,from Hadrons to Dark Matter. We thank Jim Cline and Qaisar Shafi as well as Vishnu Padmanabhan Kovilakam, Kai Schmitz, and Luca Paolo Wiggering for useful discussions.

\end{acknowledgements}

\newpage

\end{document}